\documentclass[11pt, a4paper]{article}
\usepackage[margin=1in]{geometry}
\usepackage{amsmath}
\usepackage{eufrak}
\usepackage[utf8]{inputenc}
\usepackage{enumitem}
\usepackage{subfigure}
\usepackage{bbm}
\usepackage{color}
\usepackage{bm}

\def\cl{{C_l}}

\def\tal{{{\tilde a}_{lm}}}

\def\gsim{~\rlap{$>$}{\lower 1.0ex\hbox{$\sim$}}}

\def\simpropto{\lower.2ex\hbox{$\; \buildrel \propto \over \sim \;$}}
\def\ltsim{\lower.5ex\hbox{$\; \buildrel < \over \sim \;$}}
\def\gtsim{\lower.5ex\hbox{$\; \buildrel > \over \sim \;$}}
\def\ltsim{\lower.5ex\hbox{$\; \buildrel < \over \sim \;$}}
\def\gtsim{\lower.5ex\hbox{$\; \buildrel > \over \sim \;$}}

\def\dd{\,{\rm d}}

\def\hmmpc{\ {\rm h\,Mpc^{-1}}}

\def\dd{{\rm d}}

\def\ln{{\rm ln}}

\def\pmb#1{\setbox0=\hbox{#1}%
\kern-.025em\copy0\kern-\wd0
\kern.05em\copy0\kern-\wd0
\kern-.025em\raise.0433em\box0}

\def\simlt{\lower.5ex\hbox{$\; \buildrel < \over \sim \;$}}
\def\simgt{\lower.5ex\hbox{$\; \buildrel > \over \sim \;$}}

\newcommand{\beq}{\begin{equation}}

\newcommand{\eeq}{\end{equation}}
\def\beqa{\begin{eqnarray}}
\def\eeqa{\end{eqnarray}}
\def\fixit#1{}

\def\dd{{\rm d}}

\def\cN{{\cal N}}

\usepackage{ulem}
\usepackage{natbib}
\setcitestyle{citesep={;}, aysep={,}}

\begin{document}

\centerline{\large\bf  Probing statistical isotropy of 
cosmological radio sources using 
SKA}

\bigskip
\noindent
\begin{center}
Shamik Ghosh, Pankaj Jain, Gopal Kashyap, Rahul Kothari, Sharvari Nadkarni-Ghosh

\medskip
Physics Department\\
I.I.T. Kanpur\\
Kanpur 208016, India

\medskip 
and

\medskip

Prabhakar Tiwari

\medskip
Technion- Israel Institute of Technology,
32000 Haifa, Israel
\end{center}

\bigskip
\noindent
{\bf Abstract:} 
There currently exist many observations 
which are not consistent with 
the cosmological principle. 
We review these observations with 
a particular emphasis on those relevant for Square Kilometre Array (SKA). 
In particular, several different data sets
indicate a preferred direction pointing approximately towards 
 the Virgo cluster. We also observe a hemispherical anisotropy
in the Cosmic Microwave Background Radiation (CMBR) temperature
fluctuations. Although these inconsistencies may be attributed to systematic effects, there remains the possibility that they indicate new physics and various theories have been proposed to explain them. 
One possibility, which we discuss in this review, is the generation of perturbation modes during the early pre-inflationary epoch, when the Universe may not obey the cosmological principle. 
 Better measurements will provide better 
constraints on these theories. 
In particular, we propose
measurement of the dipole in number counts, sky brightness, polarized
flux and polarization orientations of radio sources.  
We also suggest test of alignment of linear polarizations of sources 
as a function of their relative separation.
Finally we propose measurement of hemispherical anisotropy or equivalently
dipole modulation in radio sources.
\medskip


\section{Introduction}

The Big Bang model is based on the cosmological principle which states 
that the Universe is isotropic and homogeneous,
i.e. there is no preferred direction or position. It is
essentially an assumption and
cannot be proven on the basis of the symmetries of the fundamental action. 
 In particular, it applies
only in a statistical sense, after averaging over distances of order 100
Mpc. Furthermore there is a preferred frame of reference, the so called 
cosmic frame of rest. The Universe appears isotropic and homogeneous 
only in this frame. Within the Big Bang paradigm, the Universe may
not be isotropic and homogeneous at very early times. It acquires this
property during inflation. It has been explicitly shown that starting
from a wide range of anisotropic but homogeneous Bianchi models, 
the Universe quickly becomes isotropic during inflation \citep{Wald:1983}.
However other models also exist which do not obey this principle. 
In this article we review the current status of the tests of the cosmological
principle. We also review some of the theoretical attempts to explain the 
observed violations of this principle. 

Observationally it is easier to test isotropy in 
contrast to homogeneity because it requires only angular positions
of the sources. A test of homogeneity requires three dimensional
mapping of the Universe. Here we shall primarily be interested in
observations which test isotropy. However we point out that an observed
violation of isotropy may arise in a fundamental model which may
be anisotropic or inhomogeneous or both. 

Even within the Big Bang model,
the Universe is not strictly isotropic and homogeneous. It obeys 
this property only in a statistical sense in the cosmic frame 
of rest. For example, let us consider the matter density $\rho(t,\mathbf x)$
where $t$ is the cosmic time and $\mathbf x$ the comoving coordinate. 
Its spatial distribution can be expressed as,
\begin{equation}
\rho(t,\mathbf x) = \rho_0(t) + \delta\rho(t,\mathbf x)\,.
\end{equation}
Here $\rho_0(t)$ is the mean density and $\delta\rho$ the fluctuations, 
such that
\begin{equation}
\langle\delta\rho(t,\mathbf x)\rangle = 0\,.
\end{equation}
Here the angular brackets represent ensemble average. An estimate of 
this mean is obtained by averaging $\delta\rho$ over a sufficiently large
patch of the Universe. We expect this distance scale to be of order 100 
Mpc. At smaller scales the matter density shows considerable clustering
and the cosmological principle does not apply. Statistical isotropy (SI)
and homogeneity implies 
\begin{equation}
\langle\delta\rho(t,\mathbf x) \delta\rho(t,\mathbf x')\rangle = f(|\mathbf x-\mathbf x'|)\,,
\end{equation}
i.e., the two point correlations depend only on the distance between
the two points and not on the direction or the position. If we relax
the assumption of isotropy then these correlations can also depend
on the direction of the vector $\mathbf x-\mathbf x'$. If we also allow
inhomogeneity, then we can also get dependence on the mean
position $(\mathbf x+\mathbf x')/2$. As we have mentioned above, statistical
isotropy applies only in the cosmic frame of rest. If we are in 
motion with respect to this frame with velocity $\mathbf v$, then at leading
order in $|\mathbf v|$, the matter
distribution is expected to show a dipole distribution peaked in the
direction of $\mathbf v$.  

Within the Big Bang model the CMBR temperature field can
be decomposed as
\begin{equation}
T(\hat n) = T_0 + T_1\hat\lambda\cdot\hat n + \Delta T(\hat n)
\end{equation}
where 
$\hat n$ is a unit vector in the direction of observation, $T_0$
the mean temperature, $T_1$ the amplitude of the CMBR dipole, $\hat\lambda$
the dipole axis and $\Delta T$ the primordial fluctuations in temperature.
Here the dipole contains both the kinematic contribution, arising due
to local motion, as well as the contribution due to primordial fluctuations. 
Hence $\Delta T$ contains only multipoles corresponding to $l\ge 2$, 
i.e., quadrupole and higher. 
We use the spherical polar coordinates $(\theta,\phi)$ to label the 
direction of observation. 
As in the case of density fluctuations, we have
\begin{equation}
\langle\Delta T(\hat n)\rangle = 0\,.
\end{equation}
Observationally, $T_0\approx 2.73$K, $T_1/T_0\sim 10^{-3}$ and $\Delta T/T_0
\sim 10^{-5}$. 
 Statistical isotropy implies that 
the two point correlation function satisfies 
\begin{equation}
\langle\Delta T(\hat n_i) \Delta T(\hat n_j)\rangle = C(\hat n_i\cdot \hat n_j)
\,, 
\end{equation}
i.e., it is a function only of the angle between the two observation points
$\hat n_i$ and $\hat n_j$. 
It is useful to expand the temperature fluctuations in terms of the 
spherical harmonics. We obtain
\begin{equation}
{{\Delta T}(\hat n)\over T_0} = \sum_{lm}a_{lm}Y_{lm}(\hat n)
\label{eq:alm}
\end{equation}
where $a_{lm}$ are the coefficients of this expansion. These also satisfy
$\langle a_{lm}\rangle=0$. Furthermore statistical isotropy implies that
\begin{equation}
\langle{a_{lm}a^*_{l'm'}}\rangle_\text{iso} = C_{l}\delta_{ll'}\delta_{m,m'}\,,
\label{eq:corr_iso}
\end{equation}
where $C_l$ is the standard CMBR power.   

The cosmological principle is supported by the 
Cosmic Microwave Background Radiation (CMBR) and galaxy surveys. 
The observed CMBR temperature $T(\theta,\phi)$ is found to be isotropic to
a very good approximation. As mentioned above, the largest deviation from
isotropy arises due to dipole which is of order $10^{-3}$. 
The dominant dipole contribution 
arises due to the velocity of the solar system ($\mathbf v_\text{CMB}$) relative to the
cosmic frame of rest. 
Its magnitude $\mathbf{v}_\text{CMB}$ and direction $(l,b)$ in galactic coordinates are respectively found to be $369\pm 0.9$ Km/s
and  $(263.99^{\circ}\pm 0.14^{\circ},b=48.26^{\circ}\pm 0.03^{\circ})$ 
\citep{Kogut:1993,Hinshaw:2009}. The 
number density and brightness of distant radio galaxies are also observed
to be isotropic to a good approximation.  
However there are many observations which suggest a potential violation
of the cosmological principle.
In particular the local velocity $\mathbf v_\text{radio}$ extracted from the observed dipole in the number density and 
brightness of radio sources is not found to be in agreement with $\mathbf v_\text{CMB}$. The direction
agrees but the magnitude is found to be approximately three times larger.  
We review such observed violations of the cosmological principle in the next section.
In section 3 we shall present a theoretical model which may potentially explain these observations. 
In section 4 we shall discuss tests of statistical isotropy at Square Km Array 
(SKA) and will conclude
in section 5. 

\section{Observed violations of statistical isotropy}

The assumption of statistical isotropy is built into the Inflationary Big Bang model, which is the Standard Model of Cosmology. The predictions of the standard model agree remarkably well with observations which is a real success for the modern era of precision cosmology. Despite the success of the theory there are 
tantalizing evidences which highlight small but persistent departures from the predictions of the isotropic theory. Such observations are mostly in the large 
distance scale observations. In this section we will discuss some of the observed violations in statistical isotropy found in  different observations
with a particular emphasis on those relevant for SKA.

\subsection{Kinematic Dipole}
Before we discuss the major observations of SI violation, it is important to understand that the Cosmological Principle is valid only in 
the cosmic frame of rest. The Earth 
is not at rest with respect to this frame. 
It is rotating about the Sun, which in turn is rotating about the centre of the Milky Way; the Milky Way moves with respect to the Local Group barycenter, which in turn moves about the large scale structures around it. The combined motion due to these peculiar velocities ensures that our frame of observation has a relative velocity with respect to the cosmic frame of rest. This leads to a dipole
in the observer frame even if the field is isotropic in the cosmic rest frame. 
 This dipole due to Doppler shift of the CMB photons is called the kinematic dipole. 

We denote the peculiar velocity of our observation frame by $\mathbf v$ and define $\bm \beta = \mathbf v/c$. If the temperature field and direction in the cosmic rest frame are identified as $T'$ and $\hat n '$ and the unprimed symbols denote the observations in our frame, then
\begin{equation}
    T(\hat n) = \frac{T'(n')}{\gamma(1-\hat n \cdot \bm \beta)},
    \label{Eq:DopplerT}
\end{equation}
and 
\begin{equation}
    \hat n = \frac{\hat n' +[(\gamma-1)\hat n'\cdot \hat v + \gamma \beta]\hat v}{\gamma(1+\hat n' \cdot \bm \beta)}
\end{equation}
where $\gamma = \sqrt{1-\beta^2}$. Due to Doppler shifting the intensity distribution of the CMB photons gets modified. We measure $T(\hat n)$ and use
these relations to obtain the temperature field in the cosmic rest frame along with the peculiar velocity of the observation frame.

The large scale structures also acquire a dipole 
due to Doppler and aberration effects
caused by our local motion. The flux density of radio sources 
typically shows a power law dependence on frequency. Furthermore the
number density of sources depends on the flux density.  
Most large scale structure surveys operate in limited frequency ranges 
and have a lower limit on the flux density. Due to 
Doppler effect, 
the frequencies in the direction of motion of the frame are blueshifted and are redshifted in the opposite direction. Due to this effect and the  
intensity cuts on the survey, sources will shift in and out of the 
range of observations. 
Hence in the direction of motion more objects are blueshifted into the
 observation frequencies while in the other hemisphere 
more sources are redshifted out of the range. 
Combining the two effects --- the Doppler shift leads to a small dipole in a limited frequency and intensity range large scale structure survey \citep{Ellis:1984,Tiwari:2015}.

The motion of the reference frame also leads to the aberration effect. 
This produces a shift in the angular position of the source. Thus
the apparent positions of an 
 isotropic distribution of sources get  
shifted towards the direction of motion of the frame, creating a dipole.
This effect is of the same order as $\beta \sim 10^{-3}$ and is relevant for large scale structure dipole studies. Combined effect of the Doppler shift and aberration produces the \textit{kinematic dipole}.

\subsection{Observed Dipole in Large Scale Structures}
Large scale structures are essentially objects formed by non-linear physics. 
When observed on small survey volumes the non-linear physics produces structures that would deviate from isotropy and homogeneity. Thus the local non-linear components of a survey would produce a \textit{local structure dipole}. This is not a violation of SI, because the Cosmological Principle is not valid on this scale. It is only when a very large survey volume, of length scales greater than a few hundred Mpc, is considered that the Cosmological  Principle is applicable 
and can be tested for SI violations. If a dipole component is present over and above the
local structure and the kinematic dipole, then it is of cosmological origin and is called the \textit{intrinsic dipole}. We are essentially interested in the intrinsic component in SI violation study.

The most significant study of dipoles in the large scale structure has been done with the NRAO VLA Sky Survey (NVSS) radio catalogue containing 1773484 radio sources \citep{Condon:1998}. 
The survey's operating frequency is 1.4 GHz and covers the entire northern hemisphere above a declination of $-40^\circ$ and has a mean redshift $\sim 1$. 
For radio sources, in the cosmic rest frame, the flux density $S$ follows a power law relation with frequency $\nu$, $S \propto \nu^{-\alpha}$, with $\alpha \approx 0.75$. The differential number count of radio sources per unit solid angle per unit flux density follows the power law: $n(\theta, \varphi, S) \propto S^{-1-x}$, where the spectral index $x$ is close to unity. 
Due to the kinematic effects discussed above, it is clear that 
both the number counts and sky brightness would show a dipole. We denote
these by $\mathbf D_N^{\text{kin}}$ and $\mathbf D_S^{\text{kin}}$ respectively. 
 These kinematic dipoles are given by 
\begin{equation}
    \mathbf D_S^{\text{kin}}=[2+x(1+\alpha)]\bm \beta\,, \qquad \qquad \mathbf D_N^{\text{kin}}=[2+x(1+\alpha)]\bm \beta\,,
    \label{Eq:KinDipole}
\end{equation}
i.e. both are described by the same formula \citep{Ellis:1984,Tiwari:2015,Singal:2011}.  
Since the velocity of our observation frame with respect to the cosmic rest frame is already known from CMB experiments, we can make a prediction for the kinematic dipole.

The earliest attempt to extract the NVSS dipole was made by \cite{Blake:2002}, where they claimed to find the dipole
amplitude approximately two sigmas larger than the expected kinematic dipole. 
The extracted direction, however, showed good agreement with expectations.
 This was revisited later by 
several authors, who found an even larger deviation from the amplitude of
the kinematic dipole.
These results are summarised in Table \ref{Table:NVSSdipole}.
\begin{table}[h]
\centering
    \begin{tabular}{l c c c}
        \hline
        \noalign{\vskip 0.05cm} 
        Authors & $D_0 \quad (\times 10^{-2})$ & $v\text{ }  (\times 10^3 \text{  in km/s)}$ & $(l,b)$ \\
        \hline 
        \hline
        Blake \& Wall (2002) & $1.05 \pm 0.42$ & $0.9 \pm 0.3$ & $(245^\circ, 41^\circ )$   \\
        Singal  (2012) & $1.8 \pm 0.3$ & $1.32 \pm 0.54$ & $(239^\circ , 44^\circ)$  \\
        Gibelyou and Huterer (2012) & $2.7 \pm 0.5$ & $1.4 \pm 0.3$ & $(214^\circ, 15^\circ)$ \\
        Tiwari et. al. (2015) $D_N$& $1.25 \pm 0.40$ & $1.00 \pm 0.32$ & $(261^\circ, 37^\circ)$ \\ 
        Tiwari et. al. (2015) $D_S$ & $1.51 \pm 0.57$ & $1.21 \pm 0.46$ & $(269^\circ, 43^\circ)$ \\
        Rubart and Schwarz. (2013) & $1.8 \pm 0.6$ & $1.5 \pm 0.5$ & $(239^\circ, 44^\circ)$ \\
        \hline 
    \end{tabular}
    \caption{\textit{NVSS observed dipole amplitude, observation frame peculiar velocity and direction}. Collected results 
\citep{Blake:2002, Singal:2011, Gibelyou:2012, Tiwari:2015, Rubart:2013} 
for the NVSS dipole amplitude and direction with flux densities $> 20$ mJy ($>15$ mJy for Gibelyou and Huterer). Here $D_0$ is the total observed dipole and $v$ is the peculiar velocity of the observation frame, calculated from $D_0$.}
    \label{Table:NVSSdipole}
\end{table}

We note that the result obtained by \cite{Gibelyou:2012} 
shows a much larger deviation from others.  \cite{Rubart:2013} have shown that the dipole amplitude estimator used by Gibelyou and Huterer is 
biased. It has a direction bias and as a consequence their dipole direction estimates are not in agreement with  other results. The 
amplitude obtained by Blake and Wall 
 is smaller than that obtained by any of the other authors. 
Our study of the NVSS dipole \citep{Tiwari:2015} involved studying not just the number count but also the sky brightness dipole. 
Both observables show similar results with 
 amplitudes 
exceeding the kinematic dipole predictions by approximately two sigmas. 
Such excess dipole on such large 
distance scales suggests a mild signal of potential violation of SI.

The results discussed above while being intriguing need to be reassessed with other data sets due to the limitations of the NVSS catalogue. 
The NVSS is compiled by use of two different array configurations, one above declination of $-10^\circ$ and one below. This results in systematics in 
the catalogue. The mean number count becomes a function of declination. 
Plots of number count density show a large and significant dip below a declination of $-15^\circ$ and a small but linear systematic decrease with increasing right ascension. With a flux cut $>15$ mJy, the effect of these systematics can be suppressed to a level where they are no longer visible to the naked eye 
while plotting. While the work done with the NVSS data try to limit the effect of such systematics, having another deep survey with large sky coverage to test out these results would be very important before we can be sure of SI violation.

The NVSS also contains information about the polarization of the sources. It provides Stokes parameters $Q$ and $U$ for these sources. Using them we can test
the isotropy of sources with non-zero polarized flux density $P$, defined as,
 $P=\sqrt{Q^2+U^2}$. The polarized flux density, for radio sources, follows a power law, $P \propto \nu^{-\alpha_P}$, with $\alpha_P \approx 0.75$.  The differential number count per unit solid angle, per total flux density $S$ and 
polarized flux density $P$ is given as $n(\theta, \varphi, P, S) \propto S^{-1-x}P^{-1-x_P}$ in the cosmic rest frame. The kinematic dipole in the number count of significantly polarized sources and the integrated polarized flux density is given by \cite{Tiwari:2015a}:
\begin{align}
    \mathbf D_{N_P}^{\text{kin}}=[2+x(1+\alpha)+x_P(1+\alpha_P)]\bm \beta \\
    \mathbf D_P^{\text{kin}}=[2+x(1+\alpha)+x_P(1+\alpha_P)]\bm \beta
\end{align}
As in the case of Eq. \ref{Eq:KinDipole}, 
we find that these dipoles also turn out to 
be identical. 
The extracted velocities are shown in Table \ref{Table:PolDipole}. It clearly 
shows a deviation from the expectations of a kinematic dipole which may 
indicate the presence of an intrinsic dipole. 

\begin{table}[h]
\centering
    \begin{tabular}{l c c c}
        \hline
        \noalign{\vskip 0.05cm} 
        Dipole type & $D_0 \quad (\times 10^{-2})$ &  $v\text{ }  (\times 10^3 \text{  in km/s)}$ & $(l,b)$ \\
        \hline 
        \hline
        $D_{N_P}$ & $3.3 \pm 0.8$ & $2.38 \pm 0.61$ & $(207^\circ, 37^\circ )$   \\
        $D_P$ & $4.9 \pm 1.2$ & $2.87 \pm 0.68$ & $(244^\circ, 20^\circ)$ \\
        \hline 
    \end{tabular}
    \caption{\textit{NVSS dipole amplitude, observation frame peculiar velocity and direction for sources with non-zero polarized flux}. Results from \cite{Tiwari:2015a} with a lower limit on total flux density of $30$ mJy
and polarized flux density range of $0.1<P<100$ mJy.}
    \label{Table:PolDipole}
\end{table}

Study of Sloan Digital Sky Survey (SDSS) by \cite{Itoh:2009} also revealed some fascinating hints of SI violations. The SDSS 6th Data Release photometric catalogue contains over 200 million sources and covers an area of around $8000\text{ deg}^2$, with photometric data in five band passes. While the SDSS has a very high fidelity data with low and well understood systematics, its sky coverage is small --  at about 20\% with a mean redshift $\sim 0.3$. This makes the catalogue difficult to use for cosmological purpose. There are also some issues which need to be taken care of in constructing the sample for analysis. The first is to ensure that stars are carefully and reliably removed from analysis. Putting appropriate magnitude range helps in isolating the galaxies. Another well known feature of the SDSS catalogue is the presence of local clustering at large scales. The most well known feature of the SDSS is the \textit{Sloan Great Wall}, at a redshift of $\sim 0.08$. Such local clustering 
has to 
be removed reliably before the intrinsic dipole can be studied.
  The expected kinematic dipole amplitude in the SDSS is 
found to be $1.231 \times 10^{-3}$ \citep{Itoh:2009}. 
The authors worked with four galaxy samples with different ranges in brightness and photometric redshift. Of them we only discuss two here. These are those samples which are deepest, more relevant from a cosmological point of view. The results we discuss are for the bright deep (BD) and the faint deep (FD) samples. 
For both, the maximum photometric redshift is $\sim 0.9$. The authors performed a $\chi^2$ minimisation with the full covariance matrix. For the BD sample the authors obtained a dipole amplitude of $0.87^{+0.59}_{-0.57} \times 10^{-2}$ along $(l=290^\circ, b=-10^\circ) \pm 100^\circ$. The FD sample gave a dipole amplitude of $(1.21 \pm 0.23) \times 10^{-2}$ along $(l=280^\circ, b=75^\circ) \pm 33^\circ$.

The authors found a dipole excess in all but the BD sample. They suggested that possible contamination in the FD samples from incomplete star-galaxy separation and with incorrectly removed clustering in the data 
might've caused the large measured dipole in this sample. 
Another reason for difference between the two samples 
might be the small sky coverage of the survey. They hoped 
that a sky survey with a wider coverage would be able to settle the issue.

\cite{Yoon:2014} found a dipole in the Wide-field Infrared Survey Explorer-Two Micron All Sky Survey (WISE-2MASS) catalogues. The WISE catalogue has 757 million sources which are however uncategorised. The authors use the 2MASS catalogue with joint intensity limits to select data for analysis. The GAMA D2 data was used to model the redshift distribution for the WISE catalogue. The selected object field is shallow with mean redshift of 0.139 and goes up to a maximum of 0.4. They follow a method similar to that of \cite{Gibelyou:2012} to estimate the dipole. With a $20^\circ$ galactic plane cut, the result they obtained was $(5.2 \pm 0.2) \times 10^{-2}$ along $(l= 308^\circ \pm 4^\circ, b=-14^\circ \pm2 ^\circ)$, which exceeded the theoretical expectations from local structure dipole. The theoretical dipole amplitude expected being $2.3 \pm 1.2$. They did not consider the effect of the kinematic dipole which has an order of magnitude lesser contribution and could not be sufficiently tested with the shallow data. 

In the last few years the tests of SI violations with large scale structures have gathered steam. With deeper data and with greater sky coverage, better constraints can be put on SI violations and thereby constraining SI violating model parameters and mechanisms. With improvement in data fidelity and understanding of systematics, we may be able to reduce these errors and find out if truly these SI violations are consistent.

\subsection{Virgo Alignment}
A very curious feature of SI violations is the alignment of various preferred directions in different data sets. Several observations at wide range
of frequencies 
suggest a preferred direction pointing roughly towards
the Virgo supercluster, which is close to the direction of
the observed CMBR dipole. 
We have already discussed the possible presence of intrinsic dipole in
the number counts, sky brightness as well as polarized
radio flux. 
Furthermore,
the CMB quadrupole, CMB octopole, radio and optical polarizations
from distant sources also indicate a preferred direction pointing
roughly towards Virgo. Next we briefly describe each of these effects.

The distribution of polarization angles of distant radio
galaxies indicates a dipole pattern. Here the observable is
$\beta=\chi-\phi$, where $\chi$ is the linear polarization angle
and $\phi$ is the orientation angle of the galaxy. This parameter
shows a dipole distribution across the sky. The significance 
of the effect is found to be 3.5$\sigma$ after making a cut which 
eliminates the central peak in the distribution of the rotation
measures (RM) \citep{Jain:1999,Jain:2006}. The preferred direction
of the dipole is found to be $l=259^{\rm o}$, $b=62^{\rm o}$ in
galactic coordinates.

The CMBR quadrupole and the octopole, i.e. multipoles corresponding
to $l=2,3$, also indicate a preferred direction
($(l,b)\sim (250^\circ, 60^\circ)$), 
 pointing roughly towards
Virgo. Statistical isotropy would imply that these are independent of
one another as well of other multipoles, such as the dipole. 
However the preferred axis of both these multipoles
 points approximately
in the direction of the CMB dipole \citep{deOliveira-Costa:2004,Ralston:2004}. 
This is rather surprising! Furthermore, 
it is difficult to explain this alignment in terms of bias
or foreground effects \citep{Aluri:2011}.   
The procedure for extraction of the preferred
direction has been developed in \citep{deOliveira-Costa:2004,Ralston:2004,Samal:2008}. There also exist other methods for testing statistical isotropy
of CMBR \citep{Hajian:2005,Copi:2007}. 
One may either maximize the angular momentum dispersion $\langle \frac{\delta T}{T}|(\hat n \cdot \hat L)^2|\frac{\delta T}{T}\rangle$ \citep{deOliveira-Costa:2004,Bennett:2010}. 
Alternatively one may calculate the principle eigenvector of the 
power tensor for the two modes \citep{Ralston:2004,Samal:2008,Samal:2009}. 
For $l=2,3$ it has a simple interpretation. Both these multipoles appear
to be planar, i.e.,  
 all the hot and cold spots lie roughly in the plane. 
The direction
 perpendicular
to this is the preferred axis. In more detail, one finds that 
 most of the contribution to the octopole power comes from  $|m|=3$ coefficients. When maximized over direction the $|a_{3,3}|^2$ and $|a_{3,-3}|^2$ contribute approximately $94 \%$ of the total power in the octopole \citep{Bennett:2010}. This unusual planar power distribution in octopole is another CMB anomaly at large length scales.

The optical polarizations from distant
quasars show an alignment over very large distance scales 
\citep{Huts:1998,Jain:2004}, i.e. the linear polarizations of different
sources are observed to point in the same direction. 
A very strong
alignment effect is seen in the direction of Virgo as well as in
the diametrically opposite direction. 
The angular dependence of the two point
correlations of these polarizations was studied in \cite{Ralston:2004}. 
This dependence was not found to be statistically significant. 
However it is interesting that the correlations were found to  
 maximize along an axis pointing
towards Virgo \citep{Ralston:2004}. 
 Hence we see
that a wide range of phenomenon, ranging from radio number densities,
sky brightness, polarized flux, polarization angles, CMBR dipole, 
quadrupole and octopole as well as the optical polarizations from
quasars indicate a preferred direction pointing approximately towards
Virgo. Below we mention one more effect related to CMBR which also
indicates this direction.

\subsection{Dipole Modulation in CMBR}
The present era of precision cosmology was ushered in by the precision measurements of the cosmic microwave background (CMB), so our most important indicators of SI violations have come from the CMB observations. Of the various departures from SI predictions, the dipole modulation of the CMB temperature fluctuation field is the most important. The original claims were made by \cite{Hansen:2004b}, reporting a hemispherical power asymmetry in the CMB temperature observations made by the 
Wilkinson Microwave Anisotropy Probe (WMAP). The authors masked the galactic plane in the CMB temperature maps and analysed the binned angular power spectrum on circular patches of varying sizes, oriented about different directions in the sky. They reported significantly different $C_\ell$'s in the northern and southern galactic hemispheres for the multipole range $2-40$. The $2-4$ range was reported to have contribution from the galactic foreground residuals and the signal being directional along the galactic poles. The power spectrum estimates in $5-40$ range however showed asymmetry levels which could not be justified by systematics and noise. The asymmetry in $5-40$ range was found to maximize along $(57^\circ, 10^\circ)$ in Galactic coordinates, which is close to the ecliptic axis. In the frame of maximum asymmetry, they found that all the $5-40$ multipoles in the northern hemisphere have less power than than the average amplitudes, while in the southern hemisphere most of the multipoles in the range have more power than the average amplitude. The authors also claimed a similar signal of lower significance in the COsmic Background Explorer (COBE) data thereby ruling out systematics as a possible source of the signal.

\cite{Gordon:2006} proposed a model of linear modulation of the isotropic temperature fluctuation field to phenomenologically represent
hemispherical anisotropy. In this model, the temperature fluctuation $(\delta T)$ observed along a direction $\hat n$, is given by
\begin{equation}
    \delta T(\hat n) = \delta T_{\text{iso}}(\hat n)\left[1+f(\hat n)\right],
    \label{Eq:Modulation}
\end{equation} 
where $f( \hat n)$ is a direction dependent function that modulates $\delta T_{\text{iso}}$, the isotropic temperature fluctuation field\footnote{Note that we have changed the sign in front of $f(\hat n)$ from `$-$' to `$+$' to keep consistency with later work.}. The modulating function $f(\hat n)$ is assumed as $A \hat \lambda \cdot \hat n$. This linear modulation along a preferred direction $\hat \lambda$ and with amplitude $A$, would result in a dipole modulation at the surface of last scattering. However, it is important to understand that hemispherical power asymmetry is not the same as dipole modulation. A dipole modulation model will naturally give rise to hemispherical asymmetry but hemispherical power asymmetry does not necessitate a dipole modulation. 

In 2009, following the release of WMAP five-year data,
\cite{Hoftuft:2009} estimated the three parameters $A$, and two components of $\hat \lambda$ from the data, maximizing the log-likelihood for the dipole modulation model. The observed data along a direction $(\hat n)$ is written as in (\ref{Eq:Modulation}) but with an additive noise term to read $d(\hat n) = \delta T(\hat n) + N(\hat n)$.  The signal covariance matrix for such a model is given by
\citep{Hoftuft:2009}
\begin{equation}
    \textbf{S}_{\text{mod}}(\hat n, \hat m)=\left[1+A \hat \lambda \cdot \hat n\right]\textbf{S}_{\text{iso}}(\hat n, \hat m)\left[1+A \hat \lambda \cdot \hat m\right].
    \label{Eq:Covariance}
\end{equation}
The isotropic signal covariance matrix $\textbf{S}_{\text{iso}}$ is written as
\begin{equation}
    \textbf{S}_{\text{iso}}(\hat n, \hat m)=\frac{1}{4\pi}\sum_i (2\ell + 1)C_\ell P_\ell(\hat
    n \cdot \hat m)\,.
\end{equation}
Here the $P_\ell$s are the Legendre polynomials. The full covariance matrix then reads \citep{Hoftuft:2009} 
\begin{equation}
    \textbf{C}=\textbf{S}_{\text{mod}}(A,\hat \lambda)+\textbf{S}_{\text{iso}}+\textbf{N}+\textbf{F},
    \label{Eq:FullCovariance}
\end{equation}
with $\textbf{N}$ and $\textbf{F}$ as noise covariance and foregrounds respectively. Assuming the signal and noise both to be Gaussian the log-likelihood takes the form \citep{Hoftuft:2009}:
\begin{equation}
    -2 \ln \mathcal L (A, \hat \lambda) = \textbf{d}^T\textbf{C}^{-1}\textbf{d}+\ln |\textbf{C}|.
\end{equation}
The best-fit results in the $\ell \le 64$ range, obtained by maximizing 
the log-likelihood, are given in Table \ref{Table:DipoleParams}. The dipole modulation signal was claimed with a $3.3\sigma$ significance for $\ell \le 64$.

\begin{table}[h]
\centering
    \begin{tabular}{l c c }
        \hline
        Result from & A & (l,b)\\
        \hline 
        \hline
        \cite{Hoftuft:2009} (W5) & $0.072 \pm 0.022$ & $(224^\circ, -27^\circ)\pm 24^\circ$   \\
        \cite{Ade:2013is} (P13) & $0.065 \pm 0.021$ & $(226^\circ, -17^\circ)\pm 24^\circ$  \\
        \cite{Ade:2015is} (P15) & $0.066 \pm 0.021$ & $(225^\circ, -18^\circ)\pm 24^\circ$ \\ \hline
        \cite{Rath:2015} (W9) & $0.090 \pm 0.029$ & $(227^\circ, -14^\circ)$ \\
        \cite{Rath:2015} (P13) & $0.074 \pm 0.019$ & $(229^\circ, -16^\circ)$ \\
        \cite{Ghosh:2016}  (P15) & $0.078 \pm 0.019$ & $(242^\circ \pm 16^\circ, -17^\circ \pm 20^\circ)$\\ \hline 
    \end{tabular}
    \caption{\textit{Best-fit values for the dipole modulation parameters.}  W5 and W9 stand for WMAP five-year and nine-year datasets respectively, P13 and P15 stand for Planck 2013 and 2015 SMICA maps.}
    \label{Table:DipoleParams}
\end{table}

It has been shown \citep{Prunet:2005,Rath:2013} that for a dipole modulated temperature fluctuation field given by Eq. \ref{Eq:Modulation}, with the preferred direction $\hat \lambda$ chosen along $\hat z$, the two point correlation function of the spherical harmonic coefficients $a_{\ell m}$ is given by
\begin{align}
    \langle a_{\ell m}a^{*}_{\ell' m'}\rangle &= \langle a_{\ell m}a^{*}_{\ell' m'}\rangle _{\text{iso}}+\langle a_{\ell m}a^{*}_{\ell' m'}\rangle_{\text{dm}} \nonumber\\
    &= C_\ell \delta_{\ell \ell'}\delta_{mm'}+A\left(C_{\ell'}+C_\ell\right)\times \nonumber\\&\left[\sqrt{\frac{(\ell-m+1)(\ell+m+1)}{(2\ell+1)(2\ell+3)}}\delta_{\ell',\ell+1}+\sqrt{\frac{(\ell-m)(\ell+m)}{(2\ell+1)(2\ell-1)}}\delta_{\ell',\ell-1}\right]\delta_{m'm}.
    \label{Eq:Correlation}
\end{align}
This implies that for a dipole modulated temperature field, the covariance matrix, in spherical harmonic space is not diagonal. The added modulation gives rise to non-zero correlations between $\ell$ and $\ell\pm 1$  multipoles. 
So we have studied the dipole modulation feature using this property of non-zero $\ell$, $\ell+1$ correlations by defining a statistic $S_H$ as
\begin{equation}
    S_H=\sum_{\ell=\ell_{\text{min}}}^{\ell_{\text{max}}}\frac{\ell(\ell+1)}{(2\ell+1)}\sum_{m=-\ell}^{\ell}a_{\ell m}a^{*}_{\ell'm'}
    \label{Eq:Statistic}
\end{equation}
which is a summed estimate of the $\ell$, $\ell+1$ correlations in the range $\ell_{\text{min}}\le \ell \le \ell_{\text{max}}$. The analysis was performed
by setting $\ell_{\text{min}}=2$ and $\ell_{\text{max}}=64,128$ for extraction
of different parameters.  
 Some of the results of this analysis are shown in Table 
\ref{Table:DipoleParams} and show good agreement with other estimates. 

The hemispherical power asymmetry has persisted in the data for three 
generations of satellite based CMB experiments. The Planck experiment team has tested for both the hemispherical power asymmetry and dipole modulation in their CMB data, finding evidence for both \citep{Ade:2013is,Ade:2015is}. The dipole modulation signal has persisted at $\sim 3 \sigma $ level in the 2013 and 2015 data release. The results of the Planck team and the corresponding results with the  statistic $S_H$ are shown in Table \ref{Table:DipoleParams} for comparison.

A test of dipole modulation or equivalently hemispherical anisotropy
for the polarization $E$ modes has also been carried out in 
\cite{Ghosh:2016}. The low $l$ multipoles of the polarization field
are unreliable. Hence the authors only considered multipoles $l\ge 40$. 
Furthermore they did not test the significance of the effect 
since it required extensive numerical work in modelling detector noise. 
Interestingly it was found that the preferred direction in the range
$40\le l\le 100$ again points in the direction of Virgo. The 
direction starts to shift as we extend the upper limit on $l$. 
Although the statistical significance of the effect is unknown, it
is interesting that the low $l$ multipoles again prefer a direction
towards Virgo. 

\subsection{Dipole Modulation in large Scale Structures}
A signal of the dipole modulation has also been investigated in the large scale structures. The first attempt in this direction was made by \cite{Hirata:2009} using SDSS quasars. His approach to the problem of searching 
for dipole modulation in 
the large scale structures was based on the variation of the amplitude of the linear power spectrum $\sigma_8$. If the CMB hemispherical asymmetry and dipole modulation are of cosmological origins then they should be linked to the primordial curvature perturbations. Such a situation would lead to a gradient in the amplitude of the power spectrum along the preferred direction of the dipole. 
Since the growth and abundance of large scale structures is very sensitive to the value of $\sigma_8$, the gradient of this parameter can be constrained from the number variations of the large scale structures.

The SDSS quasars were chosen by Hirata to test out the variation of $\sigma_8$. This set had deep distance spread with wide angular coverage. 
Since these are SDSS objects, the systematics are fairly well understood. One of the drawbacks of the dataset chosen is that the number density of 
such quasars is small, roughly $1 \text{ deg}^{-2}$. When the preferred 
direction is fixed along that obtained by \cite{Eriksen:2007} $(l=225^\circ, b=-27^\circ)$, the amplitude of dipole modulation was found as $A=-0.0018\pm 0.0044$. 
A search for the best fit direction did not reveal a statistically significant
signal. Overall, Hirata's work is strongly indicative of no dipole modulation
in the large scale structures.

\cite{Fernandez-Cobos:2013} searched for the dipole modulation signal in the NVSS. Their approach is a logical extension of the Hoftuft et. al. method, described at the beginning of this section, to the large scale structures, working with the galaxy angular power spectra $C_\ell^{GG}$. They worked with three lower flux cuts of 2.5, 5.0 and 10.0 mJy. 
They corrected for the declination dependent systematics, only for the
case of the 2.5 mJy cut, by dividing the entire data map into 70 strips of equal area and rescaling the number density. From their simulation they forecasted a non-negligible dipole modulation with $A=0.065 \pm  0.013$ along the direction $(l=224^\circ, -14^\circ) \pm 17^\circ$. However they did not find any evidence of dipole modulation in data. The modulation amplitude $A$ was found to be $0.003\pm 0.015$ for 2.5 mJy cut, $0.011 \pm 0.016$ for 5.0 mJy cut and $0.007 \pm 0.014$ for 10.0 mJy cut, all of the amplitudes being compatible with null result.

\subsection{Alignment of linear polarizations of radio sources}
The linear polarizations of radio sources show alignment
with one another, analogous to the alignment of optical polarizations from
quasars. An alignment on the distance scale of 100 Mpc was reported
in \cite{Tiwari:2013} in the JVAS/CLASS sources with polarized flux 
greater than 1 mJy. This has subsequently been confirmed 
\citep{Shurtleff:2014,Pelgrims:2015}. An alignment on larger distance
scales for the subsample of QSOs in this data set has also been
reported in \cite{Pelgrims:2015}. 
An alignment on the scale of 100 Mpc may be expected within the
framework of Big Bang cosmology since sources show correlation
with one another on such distance scales. In \cite{Tiwari:2016} the
authors argued that this alignment is induced by the correlations
in the supercluster magnetic field. Within the framework of this model
the authors extracted the spectral index of the magnetic field on 
supercluster scales of order 100 Mpc. 
The extracted value 
was found to be equal to $2.74\pm 0.04$. 
Cosmological magneto-hydrodynamic simulations
\citep{Dolag:2002} on cluster scales of order few Mpc lead to a
spectral index of 2.70 which is, surprisingly, in 
  good agreement with the
value extracted in \cite{Tiwari:2016}. However this may be merely
a coincidence since the two refer to very different distance scales. 
The effect claimed in \cite{Tiwari:2016} needs to
be tested carefully by future surveys. The alignment might arise due
to bias and furthermore it is found that the significance of the
effect reduces considerably if the jackknife errors are taken into
account \citep{Tiwari:2016}. The authors argued that we require at least
four times larger data set in order to have a reliable confirmation
of this effect.

\subsection{Other Anomalies}
Other CMB anomalies worth mentioning are the Cold Spot and the parity asymmetry. \cite{Cruz:2004} reported an anomalous cold spot at $(l=209^\circ, b=-57^\circ)$ with a size of $10^\circ$. To understand the parity asymmetry we have to think of the temperature field being sum of even and odd parity fields. The even and odd parity can be characterised by 
\begin{align}
    P^+ & = \sum_{\ell=2}^{\ell_{\text{max}}}2^{-1}(1+(-1)^\ell)\ell(\ell+1)/2\pi C_\ell \\
    P^- & = \sum_{\ell=2}^{\ell_{\text{max}}}2^{-1}(1-(-1)^\ell)\ell(\ell+1)/2\pi C_\ell
\end{align}
The ratio $P^+/P^-$ denotes the ratio of the even parity contribution to the odd parity contribution. It was reported around 2010 \citep{Kim:2010, Aluri:2012}, that the ratio is anomalously large when summing over the largest angular scales. Summing the multipoles $2\le \ell \le 22$ the results for the ratio for WMAP 7 year data was $0.71$, indicating a larger contribution from the even parity. Both these anomalies continue to exist in the Planck CMB data.

\section{Theoretical Expectations}
It is generally believed that the effects reviewed in the previous section
are inconsistent with the Big Bang cosmological model. Although
these observations appear to be in conflict with the cosmological principle,
it has been shown in \citep{Aluri:2012a,Rath:2013a} that they can be accommodated
within the Big Bang paradigm. The basic idea is that the early 
pre-inflationary phase of the Universe may not be isotropic and 
homogeneous. It acquires this property during the early phase of
inflation. This has been explicitly demonstrated for the 
case of Bianchi models \citep{Wald:1983} which are anisotropic
but homogeneous. It 
has also been shown that, for a wide range of parameters, modes generated
during this early period can re-enter the horizon before the current
era and hence affect observations \citep{Aluri:2012a,Rath:2013a}. 
This implies that although the 
background evolution is isotropic and homogeneous, 
the perturbations need not respect
the cosmological principle. Interestingly the dominant effect is expected
for low $k$ modes, which observationally appear to show the largest deviation
from isotropy. This phenomenon has been explicitly demonstrated in
\cite{Rath:2013a} where the quadrupole and octopole alignment is
explained in terms of an early anisotropic phase of inflation. 
Similar ideas have been explored in order to explain the hemispherical
anisotropy \citep{Rath:2015,Rath:2015a,Kothari:2015,Ghosh:2016}. 
However in this case an explicit model requires either
an inhomogeneous Universe \citep{Carroll:2010,Rath:2015} or space-time noncommutativity
\citep{Rath:2015a,Kothari:2015a}. A detailed analysis of such models is
so far not available in the literature. Here we briefly review some
basic results which have been obtained by assuming a model of 
primordial 
power spectrum.

Let us first consider the primordial power spectrum in real space, 
defined as,
\begin{equation}
F(\mathbf R,\mathbf X) = \langle\delta(\mathbf x)\delta(\mathbf x^{\,\prime})\rangle
\end{equation} 
where $\delta(\mathbf x)$ is the primordial density fluctuation at
comoving coordinate $\mathbf x$, $\mathbf R=\mathbf x-\mathbf x^{\,\prime}$
and $\mathbf X = (\mathbf x+\mathbf x^{\,\prime})/2$. In \cite{Kothari:2015} the authors
consider the following inhomogeneous model:
\begin{equation}
F\left(\mathbf{R},\mathbf{X}\right)=f_{1}(R)+\sin\left(\mathbf{\lambda}\cdot\frac{\mathbf{X}}{\tau_{0}}+\delta\right)f_{2}(R),
\label{eq:Inhomo_real_space}
\end{equation}
where $\hat\lambda$ and $\delta$ are parameters and $\tau_0$ is the current
conformal time. Here the second term
represents the contribution due to inhomogeneity.  
In Fourier space, the corresponding power spectrum is given by,
\begin{equation}
\left\langle \delta(\mathbf{k})\delta^{*}(\mathbf{k}^{\prime})\right\rangle = P_{\text{iso}}(k)\delta^{3}\left(\mathbf{k}-\mathbf{k}^{\prime}\right) -\frac{i}{2}g(k_+) \left[\delta^{3}\left(\mathbf{k}-\mathbf{k}^{\prime}+\frac{\mathbf{\lambda}}{\tau_{0}}\right)-\delta^{3}\left(\mathbf{k}-\mathbf{k}^{\prime}-\frac{\mathbf{\lambda}}{\tau_{0}}\right)\right]
\label{eq:Inhomo_fourier_space}
\end{equation}
where
\[
g\left(k_{+}\right)=\int\frac{d^{3}{R}}{(2\pi)^{3}}\exp\left[i\left(\mathbf{k}+\mathbf{k}^{\prime}\right)\cdot\frac{\mathbf{R}}{2}\right]f_{2}(R),
\]
and $\mathbf k_+=(\mathbf k+\mathbf k^\prime)/2$.
This model leads to correlations between multipoles $l$ and $l\pm 1$ of
CMB, as expected in the case of dipole modulated temperature field
(see Eq. \ref{Eq:Correlation}). 
\cite{Kothari:2015} parameterize the function $g(k)$ as a power law, i.e.,
\begin{equation}
g(k) = g_0 P_{\rm iso}(k) (k\tau_0)^{-\alpha}
\end{equation} 
where $g_0$ and $\alpha$ are parameters. A fit to the CMB dipole
modulation data suggests that $\alpha\approx 1$.  
A similar analysis has also been carried out for an anisotropic but 
homogeneous model \citep{Kothari:2015}. As explained earlier, 
such a model is  
not possible in commutative spacetimes. However it may arise within
the framework of noncommutative spacetimes.

A study of the implications of such a primordial model on large scale structures
is so far not available in the literature. We expect that predictions
based on such models will become available by the time SKA becomes 
operational.

\subsection{The galaxy power spectrum }
For tests at SKA our primary aim is to study the distribution of
galaxies at large distances or equivalently their angular power spectrum
$\cl$. 
We next briefly discuss the relation between $\Lambda$CDM power spectrum 
$P(k)$ to 
$\cl$. Let $\cN(\hat r)$ be  the projected number density (per steradian) in 
the direction $\hat r$,  and $\bar \cN$  be the mean number density averaged  over the sky. We write the 
number density $\cN (\hat r)=\bar \cN(1+\Delta (\hat r))$, where $\Delta (\hat r)$ represents 
the projected number surface density contrast. 
Let the three-dimensional dark matter density contrast be represented as $\delta_m(\mathbf r,z(r))$, where ($\mathbf r,z(r)$) 
represent a unique location in space and time. The vector $\mathbf r$ stands for comoving 
distance $r$ in direction $\hat r$ and $z(r)$ is the
redshift corresponding to comoving distance $r$. Assuming linear galaxy biasing
$b(z)$ and linear growth factor $D(z)$ of density contrast we write the corresponding
galaxy contrast $\delta_g(\mathbf r,z(r)) =\delta_m(\mathbf r,z=0) D(z) b(z)$. 
Now we can write the 
theoretical expression for $\Delta (\hat r )$ as, 

\begin{eqnarray}
\label{eq:delta_th}
\Delta (\hat r) &=& \int _{0}^{\infty} \delta_g(\mathbf r, z(r)) p(r) dr  \nonumber\\
              &=&  \int _{0}^{\infty} \delta_m(\mathbf r,z=0)  D(z) b(z) p(r) dr , 
\end{eqnarray}
where  $p(r) \dd r$ is the probability of observing a galaxy between $r$ and
$(r+ dr)$. The expansion of $\Delta (\hat r)$ in spherical harmonics
and subsequent harmonic coefficients, $\tilde a_{lm}$, 
similar to equation (\ref{eq:alm}), is given as,
\begin{eqnarray}
\label{eq:alm_gal}
\tal&=&\int d \Omega \Delta(\hat r)  Y_{lm}(\hat r)\\
\nonumber &=& \int d\Omega Y_{lm}(\hat r) \int_{0}^{\infty} \delta_m(\mathbf r,z=0)  D(z) b(z) p(r) dr\; .
\end{eqnarray}

To write the harmonic coefficients, $\tilde a_{lm}$, in terms of the
$k$-space density field $\delta_{\mathbf k}$,
we expand $\delta_m(\mathbf r,z=0)$ in Fourier domain,
\begin{equation}
\delta_m(\mathbf r,z=0) =\frac{1}{(2\pi)^3}\int d^3 k \delta_{\mathbf k}{\rm e}^{i \mathbf k \cdot{  \mathbf r}}\; ,
\end{equation}
and substitute $$\label{eq:ekr}
{\rm e}^{i \mathbf k \cdot{  \mathbf r}}=4\pi \sum_{l,m} {i}^lj_l(kr) Y^*_{lm}(\hat 
{r})Y_{lm}(\hat {k})\; ,$$
where $j_l$ is the spherical Bessel function of first kind for integer $l$.
Subsequently we write
\begin{equation}
\label{eq:alm_th2}
\tal=\frac{{i}^l}{2\pi^2}\int D(z) b(z) p(r) dr \int d^3 k \delta_{\mathbf k}j_l(kr) Y^*_{lm}(\hat {k}) \; . 
\end{equation}
Following equation (\ref{eq:alm_th2}) we write the theoretical angular power spectrum ${\tilde C}_l$ as,
\begin{eqnarray}
\label{eq:clth}
{\tilde C}_l&=&\left\langle |\tal |^2 \right\rangle \nonumber\\
\nonumber &=& \frac{2}{\pi }\int dk k^2 P(k) \left\vert \int_{0}^{\infty} D(z) b(z) p(r) d r  j_l(kr)\right\vert^2 \\
          &=&   \frac{2}{\pi }\int dk k^2 P(k) W^2(k) \;.
\end{eqnarray}
where $W(k)=\int_{0}^{\infty} D(z) b(z) p(r) d r  j_l(kr)$ is the window function in $k$-space. We have also
used $\left\langle \delta_{\mathbf k}\delta_{\mathbf k'} \right\rangle=(2\pi)^3 \delta(\mathbf k-\mathbf k')P(k)$ where
$P(k)$ is $\Lambda$CDM power spectrum.

\subsection{Observational $C_l$}

The observational estimate of $C_l$ analogous to theoretical ${\tilde C}_l$ given in 
equation (\ref{eq:clth}) is, 
\begin{equation}
\label{eq:cobs}
C^{\rm obs}_l=\frac{\langle  |a^{\prime}_{lm}|^2\rangle}{J_{lm}} -\frac{1}{\bar \cN}
\end{equation}
where  $a^{\prime}_{lm} =\int_{\rm survey} d \Omega  \Delta(\hat r) Y_{lm}(\hat r)$  and 
$J_{lm}=\int_{_{\rm survey}}|Y_{lm}|^2 \dd \Omega$, the $J_{lm}$ is 
an approximate 
correction factor for the partial survey region \citep{Peebles:1980}. 
The term $\frac{1}{\bar \cN}$  removes 
the contribution from the Poissonian shot-noise.

The error in above estimate of power spectrum due to cosmic variance, sky coverage and shot-noise is as 
follows: 
\begin{equation}
\label{eq:dcl}
\Delta C_l = \sqrt{\frac{2}{(2l+1) f_{\rm sky}}} \left(C^{\rm obs}_l + \frac{1}{\bar \cN}\right)
\end{equation}
where $f_{\rm sky}$ is the fraction of sky observed in the survey. Notice that the above error 
estimate is applicable in case of the 2-point galaxy-galaxy angular power spectrum ($C^{gg}_l$). 
The lensing shear power spectrum is deduced considering shape measurements of the galaxies.  
The shear angular power spectrum error estimate is given by,  
\begin{equation}
\label{eq:dclsh}
\Delta C_l = \sqrt{\frac{2}{(2l+1) f_{\rm sky}}} \left(C^{\rm obs}_l + \frac{\sigma^{2}_{\epsilon}}{\bar \cN}\right)
\end{equation}
where $\sigma_\epsilon$ is the RMS variance of the ellipticity distribution. 
Furthermore, for the case of polarized sources, assuming that the 
polarization position angle is an unbiased tracer of the intrinsic morphological orientation of the
galaxy with a scatter of $\alpha_{\rm rms}$, the corresponding error estimate 
is as follows \citep{Brown:2011,Brown:2011dm}: 
\begin{equation}
\label{eq:dclsh1}
\Delta C_l = \sqrt{\frac{2}{(2l+1) f_{\rm sky}}} \left(C^{\rm obs}_l + \frac{ 16 \alpha^2_{\rm rms} \sigma^{2}_{\epsilon}}{\bar \cN}\right)\,.
\end{equation}

\section{Tests of statistical isotropy at SKA}
We propose the following tests of statistical isotropy in large scale 
structures:
\begin{itemize}
\item[1.] Determination of the dipole in number counts and sky
brightness of radio sources in order to test its consistency with the
kinematic dipole.
\item[2.] Determination of the dipole in number counts of
significantly polarized radio 
sources as well as in the polarized flux.
\item[3.] Testing the alignment of linear polarizations of radio sources
as a function of their relative separation.    
\item[4.] Testing the presence of dipole modulation in radio sources. 
\item[5.] Determination of the dipole anisotropy in the offsets between
linear polarization angles and the galaxy orientation angles.
\end{itemize}

\subsection{SKA technical details and capabilities}  
The SKA will be a highly flexible instrument with unprecedented observational 
capabilities. It will consist of an inner core and outer stations arranged in 
a log-spiral pattern. The full array will be extended to at least 3000 km from the 
central core.  This will be the largest radio telescope in the world and will 
revolutionize our understanding of the Universe. The SKA will operate in frequency 
range from 70 MHz  to 10 GHz (see \cite{Dewdney:2013} for more details).

The SKA  will perform both redshift (HI) and radio continuum surveys in 
the aforementioned
frequency range. There will be two phases of SKA observations. 
The final phase is expected to map out $~1$ billion galaxies over a sky 
area of $f_{\rm sky}\sim3/4$, out to a redshift of $z\sim2$. 
This will reduce the 
shot-noise in galaxy angular power spectrum (see equation (\ref{eq:cobs})) by a factor of $~3000$.
The resulting shot-noise will be $~3$ orders of magnitude lower than $\Lambda$CDM ${\tilde C}_l$ and will be 
negligible in comparison to cosmic variance (equation (\ref{eq:dcl})).

The SKA will yield measurements of various cosmological parameters with 
unmatched precision. The anisotropy tests at various scales will 
improve immensely. The dipole anisotropy observed in NVSS brightness and polarization will be clearly 
settled. At present the signal is observed at 
$\sim 3\sigma$ \citep{Tiwari:2015}. 
The radio 
galaxy biasing consideration gives similar significance for reasonable radial number density and galaxy bias values 
\citep{Tiwari:2015adi}. 
The galaxy-bias is a nuisance in relating the 
galaxy clustering to underlying dark matter distribution. The biasing is 
almost stochastic, scale-dependent, redshift dependent and 
non-linear \citep{Dekel:1999}. 
The bias determination is almost always indirect as we always need the 
underlying dark matter 
density power spectrum to extract bias from galaxy clustering.
As discussed earlier, the NVSS total source count is $\sim 1.8 \times 10^6$. 
The SKA source count is expected to be roughly two orders of magnitude
larger \citep{Wilman:2008}. This also applies to the polarized source density.  
The wide and deep polarization surveys with SKA will reach to 
$\mu$Jy flux limit.  
The deep polarization survey (2 $\mu$Jy) will probe the source population as a function of flux, luminosity 
and redshift, whereas the wide (33,000 deg$^2$, sensitive up to 10 $\mu$Jy ) survey will reveal the large scale 
clustering of polarized galaxies. 
Hence the statistical error in source counts, 
sky brightness, polarized number count as well as polarized flux
 will be sufficiently small in order to reliably extract
the signal of dipole anisotropy. However one has to carefully 
remove systematic effects from data. 

Besides the galaxy biasing described above,
the most important systematic effect is the contribution due to local clustering
dipole \citep{Blake:2002,Singal:2011,Gibelyou:2012,Rubart:2013,Tiwari:2015,Schwarz:2015}. So far this has been 
 removed by cross correlating with catalogues of
known nearby galaxies \citep{Blake:2002}. 
With SKA redshift survey the exact radial number density will be known.
The large area survey coverage and depth in redshift with SKA observation will allow us to 
measure the galaxy clustering at the
 largest scale ever. The SKA galaxy power 
spectrum will cover the turnover ($k<0.02 \hmmpc$) of $\Lambda$CDM  power spectrum.  
This will also allow a better constrain on galaxy bias. The 
NVSS survey also suffers from significant declination bias due to two 
different array configurations used for different declinations. While
this may not be an issue for SKA, a declination bias centered at the
array location may arise \citep{Tiwari:2015a}. Such a bias has been
identified in the NVSS survey, particularly for the sample with low 
flux cutoff, and can be effectively removed by the procedure described
in \citep{Tiwari:2015a}. Yet another systematic effect arises in
relating the extracted dipole from data to the local speed. The main
issue here is the deviation of the distribution of number density 
$n(S)$ as a function of the flux $S$ from a pure power law. However
it has been shown that a generalized distribution fits the data
very well and one can extract the local speed very accurately using
this fit \citep{Tiwari:2015,Tiwari:2015a}.

Further the resolved shape of billion galaxies from SKA will give the best shear measurements. 
The light rays from distant galaxies follow the geodesics, which bend according to the presence of matter 
in intervening space. This results in a shape distortion following the matter distribution 
fluctuations along the line of sight. This enables a direct mapping of mass distribution 
(luminous + non-luminous) and dark energy measurements. The statistical error in auto-shear 
power spectrum with SKA will decrease by a factor of $\sim3000$ due to high 
number surface density ($\sim10^5$ deg$^2$) and reliable shape measurements \citep{Demetroullas:2016}. 
With such huge improvement in statistics, it will be challenging to control the corresponding 
systematics. Cross-correlations between shear maps from SKA and LSST/{\it Euclid} 
can remove observational systematics.  

The enhanced polarization survey at SKA will also allow us to 
reliably test the alignment
of linear polarizations as a function of the angular separation among
galaxies \citep{Tiwari:2013,Tiwari:2016}. 
With two orders of magnitude increase in the number of sources,
the effect will be seen clearly if present in data. Furthermore  
the SKA redshift survey would allow a 3 dimensional analysis
which will provide an unambiguous test of this phenomenon, both 
at the supercluster scale \citep{Tiwari:2013,Tiwari:2016} and on larger
cosmological distance scales \citep{Pelgrims:2015}. Within the framework
of the theoretical model of \cite{Tiwari:2016}, it will allow a clean
extraction of the spectral index of the supercluster magnetic field at 
distance scales of order 100 Mpc. 
On cluster scales of order few Mpc, cosmological  
 magneto-hydrodynamic simulations lead to a spectral index of 2.7
for the corresponding magnetic field. It may be interesting to
apply the formalism proposed in \cite{Tiwari:2016} and extract the
magnetic field spectral index by studying correlations between
the radio linear polarizations at this distance scale. 
This will require large amount of
data on linear polarizations of galaxies separated by distances of order
Mpc. Such a measurement may also be feasible at SKA.

SKA will also make measurements of linear polarizations at different
frequencies for a very large sample of sources 
\citep{Beck:2004,Haverkorn:2015}. The main purpose of these observations is
the determination of Faraday rotation measures which will provide information about the
milky way magnetic field. However these will also allow measurements
of the host polarization position angles. For the case of active galaxies,
if we are also able to determine the orientation of the jets, it is
possible to test the dipole anisotropy claimed in \cite{Jain:1999}. 
We point out that extraction of rotation measures and polarization
position angles may be facilitated by the refined technique developed in
\cite{Sarala:2002}. 

\section{Discussion and Conclusions}
The tantalizing possibility that the Cosmological principle may be
violated is indicated by many observations. The most prominent
of these effects is the so called Virgo Alignment, which refers
to a wide range of phenomena indicating a preferred direction pointing
towards Virgo. The SKA has the capability to
convincingly test several of these effects.  
These include the dipole anisotropy in radio polarization
angles \citep{Jain:1999}, the dipole in the number counts and
sky brightness  
\citep{Blake:2002, Singal:2011, Gibelyou:2012, Tiwari:2015, Rubart:2013} 
and in the polarized number counts and polarized flux \citep{Tiwari:2015a}.
These observations 
may indicate that we need to go beyond the standard Big Bang cosmology.
Alternatively they may be explained by pre-inflationary 
anisotropic and/or inhomogeneous modes \citep{Aluri:2012a,Rath:2013a}.  
In either case, confirmation of this alignment effect is 
likely to revolutionize cosmology.
SKA will also test the signal of dipole modulation in large scale
structure. Finally it will test the alignment of radio polarizations.
It has been suggested that the alignment is induced by the correlations
in the cluster magnetic field \citep{Tiwari:2016}. Hence, if confirmed, this phenomenon might
provide a tool to study the statistical properties of the large
scale magnetic field. 

\section*{Acknowledgements}
Rahul Kothari sincerely acknowledges CSIR, New Delhi for the award of fellowship during the work.

\bibliographystyle{apalike}
\bibliography{references}

\begin{thebibliography}{}

\bibitem[Ade et~al., 2014]{Ade:2013is}
Ade, P. A.~R. et~al. (2014).
\newblock {Planck 2013 results. XXIII. Isotropy and statistics of the CMB}.
\newblock {\em Astron. Astrophys.}, 571:A23.

\bibitem[Ade et~al., 2015]{Ade:2015is}
Ade, P. A.~R. et~al. (2015).
\newblock {Planck 2015 results. XVI. Isotropy and statistics of the CMB}.

\bibitem[{Aluri} and {Jain}, 2012]{Aluri:2012a}
{Aluri}, P.~K. and {Jain}, P. (2012).
\newblock {Large Scale Anisotropy due to Pre-Inflationary Phase of Cosmic
  Evolution}.
\newblock {\em Modern Physics Letters A}, 27:1250014--1--1250014--11.

\bibitem[Aluri and Jain, 2012]{Aluri:2012}
Aluri, P.~K. and Jain, P. (2012).
\newblock {Parity Asymmetry in the CMBR Temperature Power Spectrum}.
\newblock {\em Mon. Not. Roy. Astron. Soc.}, 419:3378.

\bibitem[{Aluri} et~al., 2011]{Aluri:2011}
{Aluri}, P.~K., {Samal}, P.~K., {Jain}, P., and {Ralston}, J.~P. (2011).
\newblock {Effect of foregrounds on the cosmic microwave background radiation
  multipole alignment}.
\newblock {\em \mnras}, 414:1032--1046.

\bibitem[{Beck} and {Gaensler}, 2004]{Beck:2004}
{Beck}, R. and {Gaensler}, B.~M. (2004).
\newblock {Observations of magnetic fields in the Milky Way and in nearby
  galaxies with a Square Kilometre Array}.
\newblock {\em New Astronomy Reviews}, 48:1289--1304.

\bibitem[Bennett et~al., 2011]{Bennett:2010}
Bennett, C.~L. et~al. (2011).
\newblock {Seven-Year Wilkinson Microwave Anisotropy Probe (WMAP) Observations:
  Are There Cosmic Microwave Background Anomalies?}
\newblock {\em Astrophys. J. Suppl.}, 192:17.

\bibitem[Blake and Wall, 2002]{Blake:2002}
Blake, C. and Wall, J. (2002).
\newblock {Detection of the velocity dipole in the radio galaxies of the nrao
  vla sky survey}.
\newblock {\em Nature}, 416:150--152.

\bibitem[{Brown} and {Battye}, 2011a]{Brown:2011dm}
{Brown}, M.~L. and {Battye}, R.~A. (2011a).
\newblock {Mapping the Dark Matter with Polarized Radio Surveys}.
\newblock {\em \apjl}, 735:L23.

\bibitem[{Brown} and {Battye}, 2011b]{Brown:2011}
{Brown}, M.~L. and {Battye}, R.~A. (2011b).
\newblock {Polarization as an indicator of intrinsic alignment in radio weak
  lensing}.
\newblock {\em \mnras}, 410:2057--2074.

\bibitem[Carroll et~al., 2010]{Carroll:2010}
Carroll, S.~M., Tseng, C.-Y., and Wise, M.~B. (2010).
\newblock Translational invariance and the anisotropy of the cosmic microwave
  background.
\newblock {\em Phys. Rev. D}, 81:083501.

\bibitem[Condon et~al., 1998]{Condon:1998}
Condon, J.~J., Cotton, W.~D., Greisen, E.~W., Yin, Q.~F., Perley, R.~A.,
  Taylor, G.~B., and Broderick, J.~J. (1998).
\newblock {The NRAO VLA Sky Survey}.
\newblock {\em AJ}, 115(5):1693--1716.

\bibitem[{Copi} et~al., 2007]{Copi:2007}
{Copi}, C.~J., {Huterer}, D., {Schwarz}, D.~J., and {Starkman}, G.~D. (2007).
\newblock {Uncorrelated universe: Statistical anisotropy and the vanishing
  angular correlation function in WMAP years 1 3}.
\newblock {\em \prd}, 75(2):023507.

\bibitem[Cruz et~al., 2005]{Cruz:2004}
Cruz, M., Martinez-Gonzalez, E., Vielva, P., and Cayon, L. (2005).
\newblock {Detection of a non-gaussian spot in wmap}.
\newblock {\em Mon. Not. Roy. Astron. Soc.}, 356:29--40.

\bibitem[de~Oliveira-Costa et~al., 2004]{deOliveira-Costa:2004}
de~Oliveira-Costa, A., Tegmark, M., Zaldarriaga, M., and Hamilton, A. (2004).
\newblock {The Significance of the largest scale CMB fluctuations in WMAP}.
\newblock {\em Phys. Rev.}, D69:063516.

\bibitem[{Dekel} and {Lahav}, 1999]{Dekel:1999}
{Dekel}, A. and {Lahav}, O. (1999).
\newblock {Stochastic Nonlinear Galaxy Biasing}.
\newblock {\em \apj}, 520:24--34.

\bibitem[{Demetroullas} and {Brown}, 2016]{Demetroullas:2016}
{Demetroullas}, C. and {Brown}, M.~L. (2016).
\newblock {Cross-correlation cosmic shear with the SDSS and VLA FIRST surveys}.
\newblock {\em \mnras}, 456:3100--3118.

\bibitem[Dewdney et~al., 2013]{Dewdney:2013}
Dewdney, P., Turner, W., Millenaar, R., McCool, R., Lazio, J., and Cornwell, T.
  (2013).
\newblock Ska1 system baseline design.
\newblock {\em Document number SKA-TEL-SKO-DD-001 Revision}, 1(1).

\bibitem[{Dolag} et~al., 2002]{Dolag:2002}
{Dolag}, K., {Bartelmann}, M., and {Lesch}, H. (2002).
\newblock {Evolution and structure of magnetic fields in simulated galaxy
  clusters}.
\newblock {\em \aap}, 387:383--395.

\bibitem[{Ellis} and {Baldwin}, 1984]{Ellis:1984}
{Ellis}, G.~F.~R. and {Baldwin}, J.~E. (1984).
\newblock {On the Expected Anisotropy of Radio Source Counts}.
\newblock {\em MNRAS}, 206:377--381.

\bibitem[Eriksen et~al., 2007]{Eriksen:2007}
Eriksen, H.~K., Banday, A.~J., Gorski, K.~M., Hansen, F.~K., and Lilje, P.~B.
  (2007).
\newblock {Hemispherical power asymmetry in the three-year Wilkinson Microwave
  Anisotropy Probe sky maps}.
\newblock {\em Astrophys. J.}, 660:L81--L84.

\bibitem[Fern\'{a}ndez-Cobos et~al., 2014]{Fernandez-Cobos:2013}
Fern\'{a}ndez-Cobos, R., Vielva, P., Pietrobon, D., Balbi, A.,
  Mart\'{i}nez-Gonz\'{a}lez, E., and Barreiro, R.~B. (2014).
\newblock {Searching for a dipole modulation in the large-scale structure of
  the Universe}.
\newblock {\em Mon. Not. Roy. Astron. Soc.}, 441(3):2392--2397.

\bibitem[Ghosh et~al., 2016]{Ghosh:2016}
Ghosh, S., Kothari, R., Jain, P., and Rath, P.~K. (2016).
\newblock {Dipole Modulation of Cosmic Microwave Background Temperature and
  Polarization}.
\newblock {\em JCAP}, 1601(01):046.

\bibitem[Gibelyou and Huterer, 2012]{Gibelyou:2012}
Gibelyou, C. and Huterer, D. (2012).
\newblock {Dipoles in the Sky}.
\newblock {\em Mon. Not. Roy. Astron. Soc.}, 427:1994--2021.

\bibitem[Gordon, 2007]{Gordon:2006}
Gordon, C. (2007).
\newblock {Broken Isotropy from a Linear Modulation of the Primordial
  Perturbations}.
\newblock {\em Astrophys. J.}, 656:636--640.

\bibitem[{Hajian} et~al., 2005]{Hajian:2005}
{Hajian}, A., {Souradeep}, T., and {Cornish}, N. (2005).
\newblock {Statistical Isotropy of the Wilkinson Microwave Anisotropy Probe
  Data: A Bipolar Power Spectrum Analysis}.
\newblock {\em \apjl}, 618:L63--L66.

\bibitem[Hansen et~al., 2004]{Hansen:2004b}
Hansen, F.~K., Banday, A.~J., and Gorski, K.~M. (2004).
\newblock {Testing the cosmological principle of isotropy: Local power spectrum
  estimates of the WMAP data}.
\newblock {\em Mon. Not. Roy. Astron. Soc.}, 354:641--665.

\bibitem[{Haverkorn} et~al., 2015]{Haverkorn:2015}
{Haverkorn}, M., {Akahori}, T., {Carretti}, E., {Ferri{\`e}re}, K., {Frick},
  P., {Gaensler}, B., {Heald}, G., {Johnston-Hollitt}, M., {Jones}, D.,
  {Landecker}, T., {Mao}, S.~A., {Noutsos}, A., {Oppermann}, N., {Reich}, W.,
  {Robishaw}, T., {Scaife}, A., {Schnitzeler}, D., {Stepanov}, R., {Sun}, X.,
  and {Taylor}, R. (2015).
\newblock {Measuring magnetism in the Milky Way with the Square Kilometre
  Array}.
\newblock {\em Advancing Astrophysics with the Square Kilometre Array
  (AASKA14)}, page~96.

\bibitem[Hinshaw et~al., 2009]{Hinshaw:2009}
Hinshaw, G. et~al. (2009).
\newblock {Five-Year Wilkinson Microwave Anisotropy Probe (WMAP) Observations:
  Data Processing, Sky Maps, and Basic Results}.
\newblock {\em Astrophys. J. Suppl.}, 180:225--245.

\bibitem[Hirata, 2009]{Hirata:2009}
Hirata, C.~M. (2009).
\newblock {Constraints on cosmic hemispherical power anomalies from quasars}.
\newblock {\em JCAP}, 0909:011.

\bibitem[Hoftuft et~al., 2009]{Hoftuft:2009}
Hoftuft, J., Eriksen, H.~K., Banday, A.~J., Gorski, K.~M., Hansen, F.~K., and
  Lilje, P.~B. (2009).
\newblock {Increasing evidence for hemispherical power asymmetry in the
  five-year WMAP data}.
\newblock {\em Astrophys. J.}, 699:985--989.

\bibitem[{Hutsemekers}, 1998]{Huts:1998}
{Hutsemekers}, D. (1998).
\newblock {Evidence for very large-scale coherent orientations of quasar
  polarization vectors}.
\newblock {\em \aap}, 332:410--428.

\bibitem[Itoh et~al., 2010]{Itoh:2009}
Itoh, Y., Yahata, K., and Takada, M. (2010).
\newblock {A dipole anisotropy of galaxy distribution: Does the CMB rest-frame
  exist in the local universe?}
\newblock {\em Phys. Rev.}, D82:043530.

\bibitem[Jain et~al., 2004]{Jain:2004}
Jain, P., Narain, G., and Sarala, S. (2004).
\newblock {Large scale alignment of optical polarizations from distant QSOs
  using coordinate invariant statistics}.
\newblock {\em Mon. Not. Roy. Astron. Soc.}, 347:394.

\bibitem[{Jain} and {Ralston}, 1999]{Jain:1999}
{Jain}, P. and {Ralston}, J.~P. (1999).
\newblock {Anisotropy in the Propagation of Radio Polarizations from
  Cosmologically Distant Galaxies}.
\newblock {\em Modern Physics Letters A}, 14:417--432.

\bibitem[Jain and Rath, 2015]{Rath:2015a}
Jain, P. and Rath, P.~K. (2015).
\newblock {Noncommutative Geometry and the Primordial Dipolar Imaginary Power
  Spectrum}.
\newblock {\em Eur. Phys. J.}, C75:113.

\bibitem[Jain and Sarala, 2006]{Jain:2006}
Jain, P. and Sarala, S. (2006).
\newblock {Interpretation of the global anisotropy in the radio polarizations
  of cosmologically distant sources}.
\newblock {\em J. Astrophys. Astron.}, 27:443--454.

\bibitem[Kim and Naselsky, 2010]{Kim:2010}
Kim, J. and Naselsky, P. (2010).
\newblock {Anomalous parity asymmetry of the Wilkinson Microwave Anisotropy
  Probe power spectrum data at low multipoles}.
\newblock {\em Astrophys. J.}, 714:L265--L267.

\bibitem[Kogut et~al., 1993]{Kogut:1993}
Kogut, A. et~al. (1993).
\newblock {Dipole anisotropy in the COBE DMR first year sky maps}.
\newblock {\em Astrophys. J.}, 419:1.

\bibitem[Kothari et~al., 2015a]{Kothari:2015}
Kothari, R., Ghosh, S., Rath, P.~K., Kashyap, G., and Jain, P. (2015a).
\newblock {Imprint of Inhomogeneous and Anisotropic Primordial Power Spectrum
  on CMB Polarization}.

\bibitem[Kothari et~al., 2015b]{Kothari:2015a}
Kothari, R., Rath, P.~K., and Jain, P. (2015b).
\newblock {Cosmological Power Spectrum in Non-commutative Space-time}.

\bibitem[{Peebles}, 1980]{Peebles:1980}
{Peebles}, P.~J.~E. (1980).
\newblock {\em {The large-scale structure of the universe}}.

\bibitem[{Pelgrims} and {Hutsem{\'e}kers}, 2015]{Pelgrims:2015}
{Pelgrims}, V. and {Hutsem{\'e}kers}, D. (2015).
\newblock {Polarization alignments of radio quasars in JVAS/CLASS surveys}.
\newblock {\em \mnras}, 450:4161--4173.

\bibitem[{Prunet} et~al., 2005]{Prunet:2005}
{Prunet}, S., {Uzan}, J.-P., {Bernardeau}, F., and {Brunier}, T. (2005).
\newblock {Constraints on mode couplings and modulation of the CMB with WMAP
  data}.
\newblock {\em Phys. Rev. D}, 71(8):083508.

\bibitem[Ralston and Jain, 2004]{Ralston:2004}
Ralston, J.~P. and Jain, P. (2004).
\newblock {The Virgo alignment puzzle in propagation of radiation on
  cosmological scales}.
\newblock {\em Int. J. Mod. Phys.}, D13:1857--1878.

\bibitem[Rath et~al., 2015]{Rath:2015}
Rath, P.~K., Aluri, P.~K., and Jain, P. (2015).
\newblock {Relating the inhomogeneous power spectrum to the CMB hemispherical
  anisotropy}.
\newblock {\em Phys. Rev.}, D91:023515.

\bibitem[Rath and Jain, 2013]{Rath:2013}
Rath, P.~K. and Jain, P. (2013).
\newblock {Testing the Dipole Modulation Model in CMBR}.
\newblock {\em JCAP}, 1312:014.

\bibitem[Rath et~al., 2013]{Rath:2013a}
Rath, P.~K., Mudholkar, T., Jain, P., Aluri, P.~K., and Panda, S. (2013).
\newblock {Direction dependence of the power spectrum and its effect on the
  Cosmic Microwave Background Radiation}.
\newblock {\em JCAP}, 1304:007.

\bibitem[Rubart and Schwarz, 2013]{Rubart:2013}
Rubart, M. and Schwarz, D.~J. (2013).
\newblock {Cosmic radio dipole from NVSS and WENSS}.
\newblock {\em Astron. Astrophys.}, 555:A117.

\bibitem[Samal et~al., 2008]{Samal:2008}
Samal, P.~K., Saha, R., Jain, P., and Ralston, J.~P. (2008).
\newblock {Testing Isotropy of Cosmic Microwave Background Radiation}.
\newblock {\em Mon. Not. Roy. Astron. Soc.}, 385:1718.

\bibitem[Samal et~al., 2009]{Samal:2009}
Samal, P.~K., Saha, R., Jain, P., and Ralston, J.~P. (2009).
\newblock {Signals of Statistical Anisotropy in WMAP Foreground-Cleaned Maps}.
\newblock {\em Mon. Not. Roy. Astron. Soc.}, 396:511.

\bibitem[{Sarala} and {Jain}, 2002]{Sarala:2002}
{Sarala}, S. and {Jain}, P. (2002).
\newblock {A Circular Statistical Method for Extracting Rotation Measures}.
\newblock {\em Journal of Astrophysics and Astronomy}, 23:137.

\bibitem[Schwarz et~al., 2015]{Schwarz:2015}
Schwarz, D.~J., Bacon, D., Chen, S., Clarkson, C., Huterer, D., Kunz, M.,
  Maartens, R., Raccanelli, A., Rubart, M., and Starck, J.-L. (2015).
\newblock {Testing foundations of modern cosmology with SKA all-sky surveys}.
\newblock {\em PoS}, AASKA14:032.

\bibitem[Shurtleff, 2014]{Shurtleff:2014}
Shurtleff, R. (2014).
\newblock {Testing the Alignment Tendency of Some Polarized Radio Sources}.
\newblock {\em arXiv:1408.2514}.

\bibitem[Singal, 2011]{Singal:2011}
Singal, A.~K. (2011).
\newblock {Large peculiar motion of the solar system from the dipole anisotropy
  in sky brightness due to distant radio sources}.
\newblock {\em Astrophys. J.}, 742:L23.

\bibitem[Tiwari and Jain, 2013]{Tiwari:2013}
Tiwari, P. and Jain, P. (2013).
\newblock {Polarization Alignment in JVAS/CLASS flat spectrum radio surveys}.
\newblock {\em Int. J. Mod. Phys.}, D22(14):1350089.

\bibitem[Tiwari and Jain, 2015a]{Tiwari:2015a}
Tiwari, P. and Jain, P. (2015a).
\newblock {Dipole Anisotropy in Integrated Linearly Polarized Flux Density in
  NVSS Data}.
\newblock {\em MNRAS}, 447:2658--2670.

\bibitem[Tiwari and Jain, 2015b]{Tiwari:2016}
Tiwari, P. and Jain, P. (2015b).
\newblock {Extracting Spectral Index of Intergalactic Magnetic Field from Radio
  Polarizations}.
\newblock {\em ArXiv e-prints}.

\bibitem[Tiwari et~al., 2015]{Tiwari:2015}
Tiwari, P., Kothari, R., Naskar, A., Nadkarni-Ghosh, S., and Jain, P. (2015).
\newblock {Dipole anisotropy in sky brightness and source count distribution in
  radio NVSS data}.
\newblock {\em Astropart. Phys.}, 61:1--11.

\bibitem[Tiwari and Nusser, 2015]{Tiwari:2015adi}
Tiwari, P. and Nusser, A. (2015).
\newblock {Revisiting the NVSS number count dipole}.
\newblock {\em arXiv:1509.02532}.

\bibitem[{Wald}, 1983]{Wald:1983}
{Wald}, R.~M. (1983).
\newblock {Asymptotic behavior of homogeneous cosmological models in the
  presence of a positive cosmological constant}.
\newblock {\em \prd}, 28:2118--2120.

\bibitem[Wilman et~al., 2008]{Wilman:2008}
Wilman, R.~J., Miller, L., Jarvis, M.~J., Mauch, T., Levrier, F., Abdalla,
  F.~B., Rawlings, S., Klöckner, H.-R., Obreschkow, D., Olteanu, D., and
  Young, S. (2008).
\newblock A semi-empirical simulation of the extragalactic radio continuum sky
  for next generation radio telescopes.
\newblock {\em Monthly Notices of the Royal Astronomical Society},
  388(3):1335--1348.

\bibitem[Yoon et~al., 2014]{Yoon:2014}
Yoon, M., Huterer, D., Gibelyou, C., Kovács, A., and Szapudi, I. (2014).
\newblock {Dipolar modulation in number counts of WISE-2MASS sources}.
\newblock {\em Mon. Not. Roy. Astron. Soc.}, 445:L60--L64.

\end{thebibliography}

\end{document}